\documentclass[aps,prl,nofootinbib]{revtex4}

\usepackage{graphicx}
\usepackage{amsfonts}
\usepackage{amsmath}
\newcommand{\be}{\begin{eqnarray}}
\newcommand{\ee}{\end{eqnarray}}

\begin{document}


\title{Dominant folding pathways of a peptide chain,  from \emph{ab-initio} quantum-mechanical simulations}

\author{Silvio a Beccara}
\affiliation{Dipartimento di Fisica  Universit\`a degli Studi di Trento, Via Sommarive 14, Povo (Trento), I-38050 Italy.}
\affiliation{INFN, Gruppo Collegato di Trento, Via Sommarive 14, Povo (Trento), I-38123 Italy.} 
\author{Pietro Faccioli~\footnote{Corresponding author: faccioli@science.unitn.it}} 
\affiliation{Dipartimento di Fisica  Universit\`a degli Studi di Trento, Via Sommarive 14, Povo (Trento), I-38050 Italy.}
\affiliation{INFN, Gruppo Collegato di Trento, Via Sommarive 14, Povo (Trento), I-38123 Italy.} 
\author{Giovanni Garberoglio}
\affiliation{Interdisciplinary Laboratory for Computational Science (LISC), FBK-CMM and University of Trento, 
Via Sommarive 18, I-38123 Povo, Trento, Italy}, 
\author{Marcello Sega}
\affiliation{Dipartimento di Fisica  Universit\`a degli Studi di Trento, Via Sommarive 14, Povo (Trento), I-38050 Italy.}
\affiliation{INFN, Gruppo Collegato di Trento, Via Sommarive 14, Povo (Trento), I-38123 Italy.} 
\author{Francesco Pederiva}
\affiliation{Dipartimento di Fisica  Universit\`a degli Studi di Trento, Via Sommarive 14, Povo (Trento), I-38050 Italy.}
\affiliation{INFN, Gruppo Collegato di Trento, Via Sommarive 14, Povo (Trento), I-38123 Italy.} 
\author{Henri Orland}
\affiliation{Institut de Physique Th\'eorique, Centre d'Etudes de Saclay, CEA, IPhT, F-91191, Gif-sur-Yvette, France.}

\begin{abstract}
Using the Dominant Reaction Pathways (DRP) method, we perform an \emph{ab-initio} quantum-mechanical simulation of a conformational transition  of a peptide chain. The method we propose makes it possible to investigate the out-of-equilibrium dynamics of these systems, without resorting to an empirical representation of the molecular force field. It also allows to study rare transitions involving rearrangements in the electronic structure.
By comparing the results of the \emph{ab-initio} simulation with those obtained employing a standard force field, we discuss its capability to describe the non-equilibrium dynamics of conformational transitions. 
\end{abstract}

\maketitle


\section{Introduction}

  The theoretical investigation of conformational transitions of polypeptide chains is usually performed using techniques like molecular dynamics (MD),  Monte Carlo, transition path sampling based on classical Molecular Mechanics (MM) approaches\cite{mol_modeling}. MM methods are computationally very efficient, thus making simulations of molecules with thousands of atoms feasible on modern computer clusters\cite{foldingathome,anton}. In many cases the outcome of the simulations compares favorably with experimental results \cite{ponder03}.

MM methods are based on an empirical representation of the potential energy function of molecules (the so--called force field) which relies on a chemical model of the bonding fitted on quantum calculations and experimental results. A force field is defined by the functional form of the different components that make it up, and by the values of a set of parameters appearing in the components. The parameters are usually determined based on the equilibrium configuration of molecules. This approach is acceptable when the focus is on small thermal oscillations, but may become inadequate when the system undergoes out-of-equilibrium transitions. 

Another limitation of the MM approach arises when the transition involves a rearrangement of the electronic structure, as is the case for the cleavage or formation of chemical bonds, like for instance a sulphur bridge.

All these problems could in principle be solved by adopting a quantum mechanical approach to the dynamics of the molecule. Given the formidable complexity of a full quantum description, two approximations are usually invoked in \textit{ab-initio} simulations: the Born-Oppenheimer separation of the dynamics of nuclei and electrons\cite{born-oppenheimer}, and a classical treatment of nuclear degrees of freedom.

However, for molecular systems the size of poly-peptide chains, the quantum calculation of the molecular energy of a conformation is computationally quite expensive. 
As a result, the \emph{ab-initio} quantum-mechanical approach is usually adopted to infer static properties or to study the dynamics over very short time intervals, typically up to hundreds of picoseconds \cite{dalperaro}.  On the other hand, the  time scales involved in the conformational transitions of poly-peptide  chains range from nanoseconds ---for the rotations around dihedral angles--- to milliseconds or even seconds ---for the formation of tertiary structures in proteins---. 
 
In this work, we show that the DRP formalism~\cite{DRP1,DRP2,DRPtest1,DRPtest2,DRPtest3,Elber2} provides
a rigorous and computationally efficient framework which makes it
feasible to investigate \emph{ab-initio} the folding dynamics of 
peptide chains.

The DRP approach is a  method which yields the 
statistically most significant reaction pathways, in systems described by the
over-damped Langevin equation. Its computational advantage resides in
the fact that it does not waste CPU time in simulating thermal
oscillations in (meta)-stable states visited during the reaction
and that the sampling of the transition pathways is performed at
constant spatial displacement steps, rather than at constant time
steps. 
The reliability of the DRP approach when applied to investigating
thermally activated conformational reactions of peptide chains
within an MM framework has been tested in a series of works based on both
atomistic\cite{DRPtest1} and coarse-grained
models~\cite{DRPtest2,DRPtest3}. In particular, in \cite{DRPtest3} the
folding trajectories obtained by means of Molecular Dynamics (MD)
simulations were compared directly with those calculated in the DRP
approach, and the two methods were found to give consistent results.

In \cite{QDRP} the DRP formalism was applied to perform an
\emph{ab-initio} calculation of the dominant pathways in the
cyclobutene $\rightarrow$ butadiene transition, a thermally activated reaction
which involves the breaking and formation of covalent bonds.  In this approach,
the electronic structure and the molecular energy of the chain was
determined at each step of the calculation, by approximatively solving 
the Schr\"odinger equation.

\begin{figure*}
 \begin{center}
 \includegraphics[width=10cm]{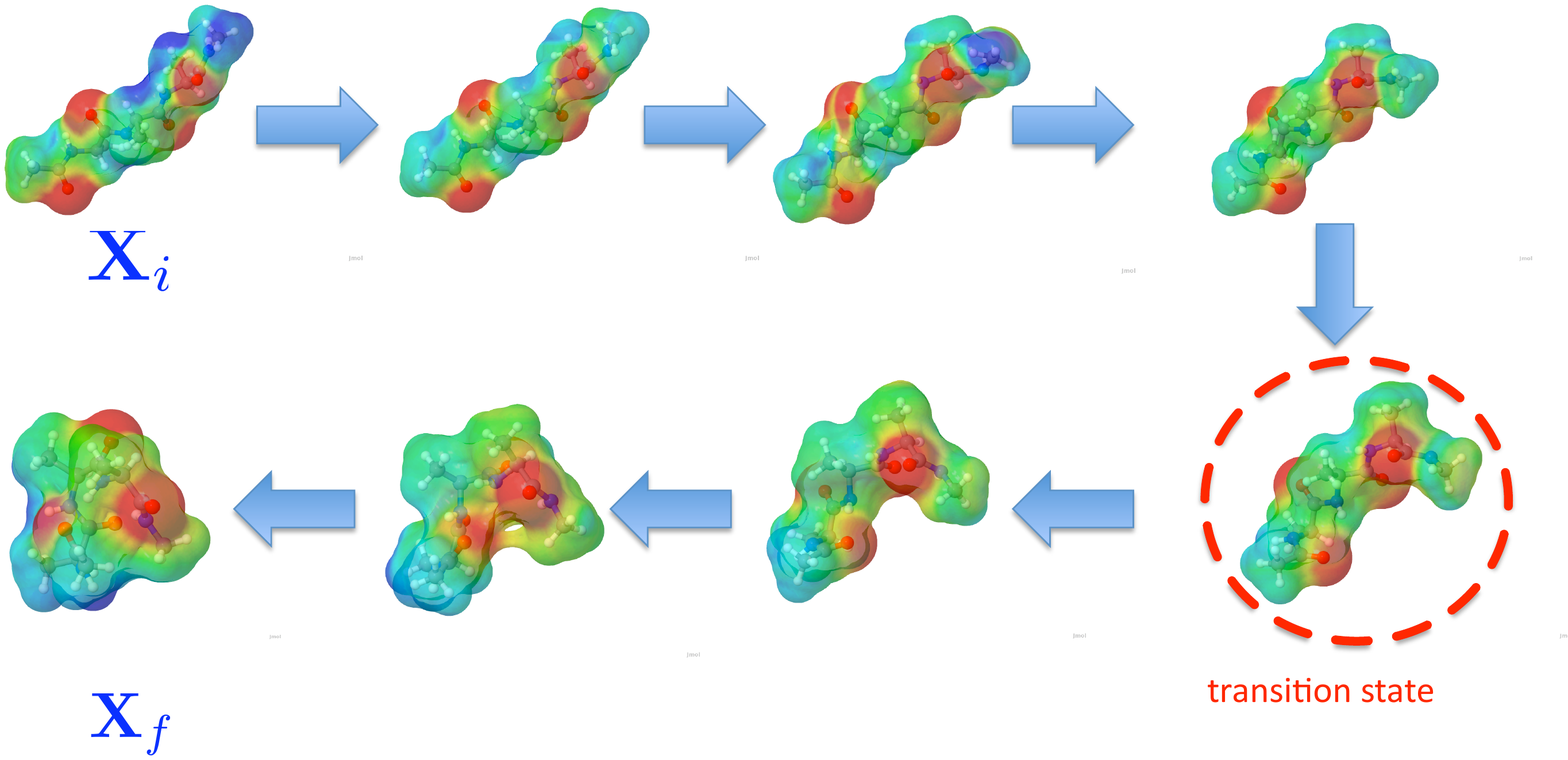}
 \caption{Configurations on the dominant reaction pathway calculated \emph{ab-initio}. The colors on the surface represent the projection of the molecular electro-static potential on the solvent accessible surface. }
 \label{frames}
 \end{center}
 \end{figure*}

In the present work, we use the same approach to compute
\emph{ab-initio} the dominant reaction pathway for a conformational
transition of tetra-alanine, leading to the formation of the
elementary unit of a helix --- see
Fig. \ref{frames}---.  

Our first goal is to show that by using the DRP formalism the \emph{ab-initio} simulation of the
reaction can be performed at a relatively low computational
cost. Indeed, the calculation of the most probable pathway connecting
a single initial and final configuration required about 23,000 CPU
hours.

A second goal of the present work is to address the question whether
force fields which are fitted on \emph{ab-initio} calculations 
of equilibrium properties also agree with quantum mechanical calculations 
in non-equilibrium conditions. To this goal, we compare
the dominant reaction pathways obtained \emph{ab-initio} with those
calculated with the AMBER--99 force field \cite{ponder03}. We find that the
classical approximation describes with reasonable accuracy the dynamics also in the
transition region, producing trajectories which are in
semi-quantitative agreement with those obtained \emph{ab-initio}.

Finally, as an example of an observable which cannot be computed from
classical MM simulations, we analyze the evolution of the partial
charges of the atoms which are involved in the hydrogen bonds in the helix
configuration. 

 \section{Model}
\subsection{The ab-initio dominant reaction pathways approach}
 
We consider a generic molecule consisting of  $N$ atoms with nuclear coordinates ${\bf X}\equiv ({\bf x}_1,\ldots, {\bf x}_N)$,  in contact with a thermal-bath at temperature $T$. The  atomic nuclei are assumed to be classical point-like particles, evolving according to Langevin dynamics.  The electrons are coupled quantum mechanically to the nuclear positions, in the Born--Oppenheimer approximation, i.e. their wave-function is assumed to instantaneously relax to the ground state, for each nuclear configuration ${\bf X}$.  

On time-scales larger than a ps the dynamics of atoms in
proteins is well described by the over-damped limit of the Langevin
equation
\begin{equation} \label{langev_eq}
 \dot {\bf X}   = -\frac{D}{k_B T} {\bf \nabla}U({\bf X}) + {\bf \eta}(t).
\end{equation} 

In this Eq., $k_B$ is the Boltzmann constant, $T$ is the temperature
and $D$ is the diffusion coefficient, which we shall assume to be the
same for all atoms (the generalization to the case in
which each atom has a different diffusion coefficient is
straightforward).  $U({\bf X})$ is the molecular energy for system
with the nuclei in configuration ${\bf X}$.  In MM 
simulations this quantity is given by the force field. In \emph{ab-initio} simulations, $U({\bf X})$
is the sum of the ground-state energy of the electron wave-function
and of the electrostatic energy of the classical nuclei.

In Eq. (\ref{langev_eq}), ${\bf \eta}(t)$ is a random noise with Gaussian distribution, zero average and correlation given by 
$$
\langle \eta_i(t) \eta_j(t') \rangle = 2~D\ \delta_{ij} \ \delta(t-t'),
$$ 
where $i, j$ label all the nuclear degrees of freedom. 

Let us now consider the probability for the molecule  to be found in the configuration ${\bf X}_f$ at time $t_f$, provided it was prepared in some initial configuration ${\bf X}_i$ at time $t_i$. 
Such a conditional probability can be represented in the following path integral form ---for the details of the derivation see e.g. \cite{DRP2}---:
\begin{eqnarray}
\label{ppath}
 P({\bf X}_f, t_f | {\bf X}_i, t_i ) = e^{-\frac{U({\bf X}_f) - U( {\bf X}_i ) }{2k_BT} }
\int_{{\bf X}_i}^{{\bf X}_f} \mathcal{D} {\bf X}(\tau) \ e^{-S_{eff} [ \mathbf{X}(\tau)]},
\end{eqnarray} 
where $S_{eff}[{\bf X}(\tau)]$ is called the effective action functional calculated on the trajectory ${\bf X}(\tau)$ and is given by
\begin{eqnarray} \label{effact}
 S_{eff}[{ \bf X}(\tau)] = \int_{t_i}^{t_f}d\tau~\left( \frac {1} {4~D} ~\dot{\bf X}^2(\tau)+ V_{eff}[ {\bf X}(\tau)] \right).
\end{eqnarray}
$V_{eff} ({\bf X})$ is called the effective potential, and reads
\begin{equation} \label{eff_pot}
 V_{eff} ({\bf X}) = \frac{D}{4 ~(k_B T)^2} \left[ \left| {\bf \nabla}U({\bf X})  \right| ^2 - 2~k_B
T\; \nabla^2U({\bf X}) \ \right].
\end{equation} 
Notice that, if ${\bf X}_f$ and ${\bf X}_i$ are product and reactant configurations respectively, then the factor $\exp(- S_{eff}[{\bf X}])$ inside the path integral expression (\ref{ppath}) 
represents the statistical weight of a given reactive trajectory ${\bf X}(t)$.
 
The most probable (or  dominant) reaction pathways are those which minimize the effective action functional $S_{eff}[{\bf X}]$. 
Hence, they are  the solutions of the classical equations of motion generated by the effective action, i.e. 
\be
\ddot {\bf X} = 2~D ~\nabla V_{eff}({\bf X}),  
\label{eom}
\ee
with boundary conditions ${\bf X}(t_i)={\bf X}_i$ and ${\bf X}(t_f)={\bf X}_f$.

Note that the dynamics described by the equation of motion (\ref{eom}) conserves the effective energy $E_{eff}= \frac {1}{4 D}  \dot { {\mathbf X}}^2(t)  - V_{eff} \left( {\mathbf X}(t) \right). $ 
This property allows  to switch from the {\it time}-dependent Newtonian description to the equivalent  {\it energy}-dependent Hamilton--Jacobi (HJ) description. 
In the HJ framework, the most probable pathways connecting the given initial and final configurations can be shown to be those which minimize the target HJ~functional
\be
 S_{HJ}[{\bf X} ]= \int_{{\bf X}_i}^{{\bf X}_f} dl \; \sqrt{ \frac{1}{D} \left[ E_{eff} + V_{eff}\left({\bf X}(l) \right) \right]},
 \label{hj-action}
\ee 
where $dl = \sqrt{d {\bf X}^2}$ is the  elementary   displacement in configuration space,  along the dominant reaction path. 
Note that the dominant trajectory in Eq.~(\ref{hj-action}) is parametrized in terms of the curvilinear abscissa $l$, which plays the role of  the reaction coordinate and measures the total distance covered along the reaction pathway. 

The computational difficulty of investigating thermally activated
transitions by ordinary MD simulations is related to the decoupling of
the time scales characterizing the dynamics of the system. The
computational advantage of the DRP approach with respect to MD
simulations comes from the fact that, by switching to the HJ
formulation, the time variable $t$ has been replaced by the
curvilinear abscissa $l$. The key point is that molecular systems are
\emph{not} characterized by a decoupling of the intrinsic length
scales. As a result, typically only $N_s=10-100$ space displacement
steps are sufficient to attain a realistic representation of the
path. This number should be compared to the $10^9-10^{12}$ MD
time steps required to simulate a single transition with
mean-first-passage time in the $\mu$s --- ms range.  Ultimately, such
a huge computational gain originates from the fact that the DRP does
not waste time to simulate the dynamics of the system when it is trapped
in metastable states.

Although in the HJ formulation the time variable has been replaced by the curvilinear abscissa $l$, the DRP formalism retains  information about the time evolution of the system. Indeed,  the time $t \left[{\bf X} \right] $ at which the configuration ${\bf X}$ is visited, during the most probable reaction pathway is given by 
\begin{eqnarray}
t \left( {\bf X} \right) = \int_{{\bf X}_i}^{{\bf X}} dl \frac{1}{\sqrt{ 4~D~\left[ E_{eff}+V_{eff}\left( {\bf X}(l) \right) \right]}}.
\label{time}
 \end{eqnarray}
From this equation it follows that the choice of the effective energy parameter $E_{eff}$ determines the total time of the transition. 
In particular, the longest possible transition path time is obtained  by choosing $E_{eff}=-V_{eff}({\bf X}_f)$~
\cite{DRP2, DRPtest2}, which is  a positive number if ${\bf X}_f$ is an equilibrium configuration.  
We stress the fact that the total time $t_{tot}= t({\bf X}_f)$ is much shorter than the mean-first-passage time, as it corresponds to the time it takes to reach the product, {\it once the system has left the reactant state}. 

Once  the dominant path has been determined, it is also possible to identify the configuration ${\bf X}_{ts}$ which belongs to the transition state, 
defined  in terms of commitment analysis.
~This is achieved by requiring  that the probability in the saddle-point approximation
to diffuse back  to the initial configuration ${\bf X}_i$,  $P({\bf X}_i, t_f|{\bf X}_{ts},t_i)$
 equates that of evolving toward the final configuration ${\bf
X}_f$, $P({\bf X}_f, t_f|{\bf X}_{ts}, t_i)$.  In the saddle-point
approximation, this condition leads to the simple equation
\cite{DRPtest1}: \be \frac{U({\bf X}_f)-U({\bf X}_i)} {2 k_B T } =
S_{HJ}([{\bf X}(l)];{\bf X}_{ts},{\bf X}_i)-S_{HJ}([{\bf X}(l)];{\bf
X}_{ts},{\bf X}_f).  \label{transition} \ee 

\subsection{Details of the simulations}
 
We have studied the folding of tetra-alanine at a temperature $T=300 K$. We discretized the path using $N_s=16$ equal displacement slices. The effective energy parameter $E_{eff}$ was chosen to be slightly larger than the value corresponding to the longest possible transition time, i.e. 
\be
E_{eff} =  - \frac{3}{2}~V_{eff}({\bf X}_f).
\ee

The molecular energy in the quantum simulations, $U({\bf X})$, was evaluated by solving the  Schr\"odinger Eq. in the Parameterized Model~3~(PM3) scheme, in the MOPAC implementation\cite{PM3}. The choice of a semi-empirical quantum mechanical method was made because it combines a very low computational cost with a reasonable description of the hydrogen bond energetics~\cite{PM3test1, PM3test2}. In addition,  parameterized model quantum mechanical calculations have been recently shown to reliably describe various 
types of non-covalent complexes~\cite{PM6test}. 

Since one of the purposes of this work is to compare classical and quantum descriptions of the inter-atomic interactions,  we chose to neglect solvation terms in both the MM and \emph{ab-initio} simulations. However, the inclusion of such contributions at the implicit level in the quantum-mechanical simulations does not lead to a significant increase of the computational cost, as the bottleneck of the calculation is the solution of the Schr\"odinger Eq.,  at each step of the minimization. 
 
Finding the dominant reaction pathway amounts to minimizing a discretized version of the effective HJ functional:
\begin{equation} \label{effact_discr}
 S_{HJ}^{d} [{\bf X}]= \sum_{i=1}^{N_s-1} \sqrt{ \frac{1}{D} \left[
E_{eff} + V_{eff}\left( {\bf X}_i \right) \right] } \; \Delta l_{i,
i+1}
\end{equation}
where  the effective potential $V_{eff}({\bf X})$ is determined according to (\ref{eff_pot})  by numerically differentiating the molecular potential energy $U({\bf X})$. $\Delta l_{i, i+1}$ is the Euclidean distance between the slices $i$ and $i+1$, i.e
$
  \Delta l_{i, i+1} = \sqrt{ \left| {\bf X}_{i+1} - {\bf X}_{i} \right| ^2 }.
$

In the discretized representation of the HJ effective action
(\ref{effact_discr}), the width of the distribution of the Euclidean
distances between consecutive path slices, $\Delta l_{i,i+1}$, should
not be allowed to increase in an uncontrolled way, in order to prevent
all frames to collapse into the reactant or product configurations. As
discussed in \cite{QDRP}, the most convenient way to achieve this is
to introduce a Lagrange multiplier, which holds fixed at $0.2$ the
ratio between the mean-square deviation from the average of the
inter-slice distances $\sigma^2$ 
of the average square inter-slice distance $\langle \Delta l^2 \rangle$.

The global minimization of the HJ effective action  (\ref{effact_discr}) is in general a very challenging task. The main difficulties arise from the  ruggedness  of the effective potential and the high dimensionality of the system. Indeed, most commonly used global optimization algorithms --- such as e.g. simulated annealing--- tend to get stuck in secondary minima of the action functional. On the other hand, the results of a  DRP calculation can be considered reliable only if  the minimization algorithm explores a significant region of the path space. In fact,  in the opposite scenario the calculated dominant paths would be strongly biased by the choice of the initial trial path. 

In our previous work \cite{QDRP} we tested the performances of several global minimization algorithms, and we found that the Fast Inertial Relaxation Engine (FIRE)~\cite{FIRE} method was performing best. The minimization protocol which was adopted in the present work was  the following: we started from a high-temperature (800 K) MD trial unfolding path, from the helix configuration. We then relaxed the path by means of a Nudged Elastic Band (NEB) \cite{NEB} minimization, followed by a zero-temperature DRP minimization, and finally by a finite temperature DRP minimization.\\

\section{Results}
A sequence of  configurations which are visited by the dominant reaction pathway  determined from \emph{ab-initio} calculations is shown in Fig. \ref{frames}. The dashed circle highlights the configuration along the path which is representative of the transition state,  according to Eq.  (\ref{transition}). 

Polyalanine chains form $\alpha$-helices, stabilized by a sequence of $i-i+4$ hydrogen bonds. On the other hand,  in the  tetra-alanine molecule there are significant termination effects, and the minimum energy configuration is slightly distorted from that of an ideal $\alpha$-helix. In particular, the hydrogen bonds stabilizing the tetra-alanine  system occur between the  $O-6$ and the $H-28$ atoms and the $O-16$ and the $H-38$ atoms.

\begin{figure*}
 \begin{center}
 \includegraphics[width=8cm]{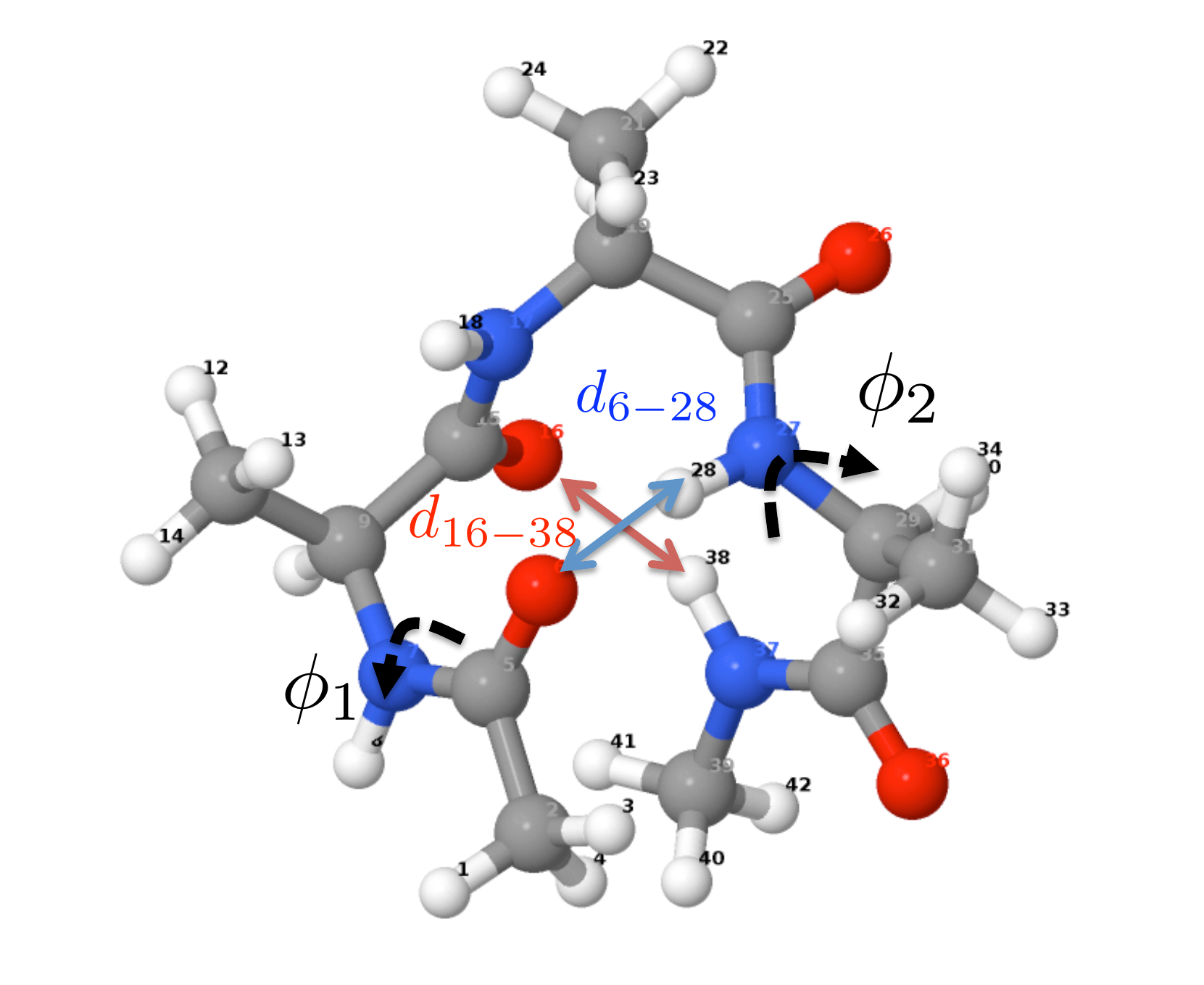}
 \end{center}
 \caption{The definition of the order parameters $d_{6-28}$, $d_{16-38}$, $\phi_1$ and $\phi_2$ used to analyze the dominant reaction pathways.}
 \label{figorderparameters} 
 \end{figure*}

In order to perform a quantitative analysis of the transition, let us study the evolution along the dominant path of the following  order parameters ---see Fig. \ref{figorderparameters}---:
\begin{itemize}
\item the distance $d_{6-28}$ between the $O-6$ and the $H-28$ atoms
\item the distance $d_{16-38}$ between the $O-16$ and the $H-38$ atoms
\item the dihedral angle $\phi_1$ between the atoms $C-5, N-7, C-9, C-15$ 
\item the dihedral angle $\phi_2$ between the atoms $C-25,N-27,C-29, C-35$.  
\end{itemize}

\begin{figure*}
 \includegraphics[width=14 cm]{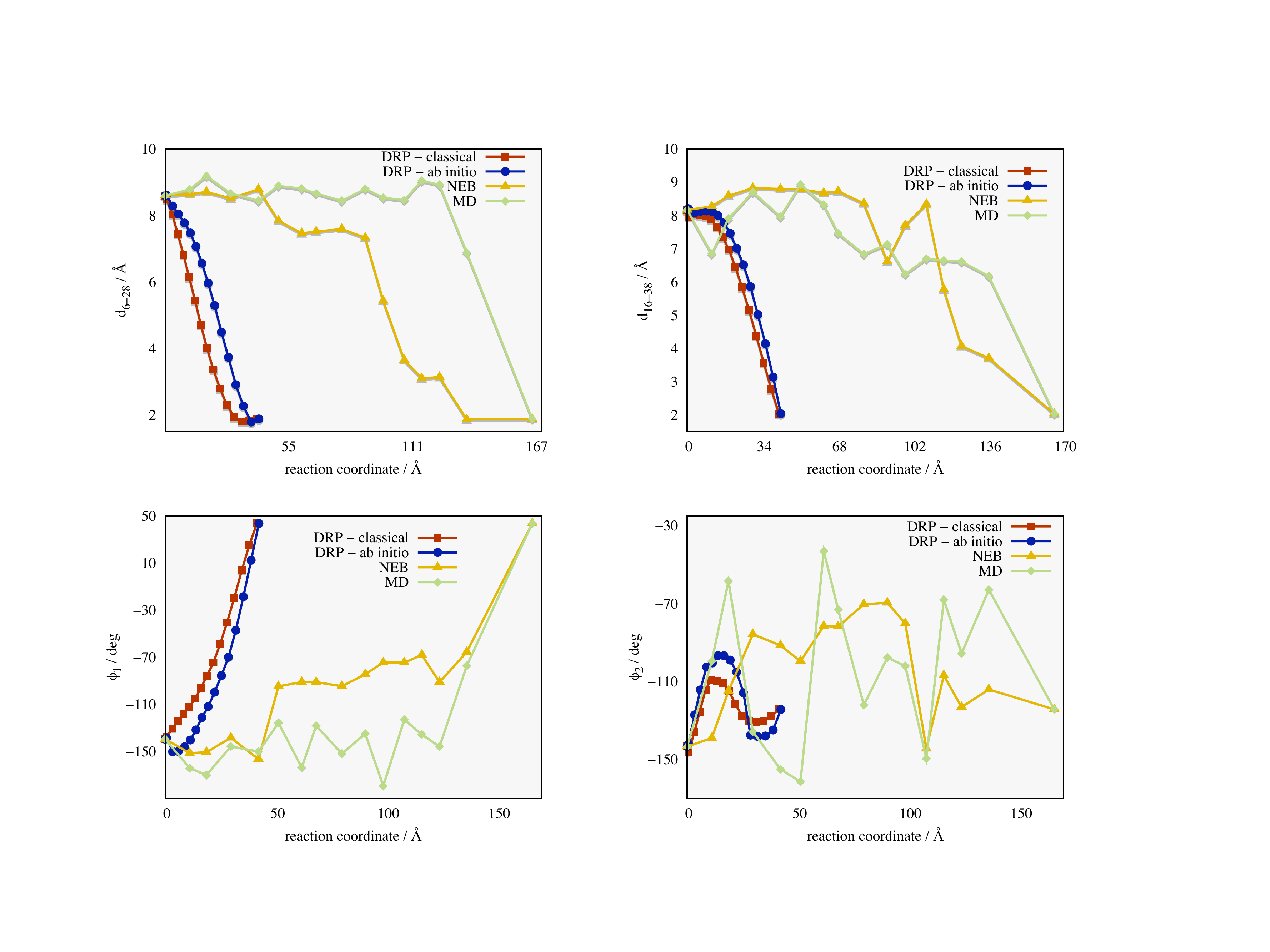}
 \caption{Upper-left panel: Evolution of the order parameter $d_{6-28}$ as a function of the reaction coordinate $l$ for  four different paths.
 Upper-right panel: Evolution of the order parameter $d_{16-38}$ as a function of the reaction coordinate $l$ for  four different paths.
 Lower-left panel: Evolution of the order parameter $\phi_{1}$ as a function of the reaction coordinate $l$ for  four different paths.
 Lower-right panel: Evolution of the order parameter $\phi_2$ as a function of the reaction coordinate $l$ for  four different paths.}
 \label{orderparameters}. 
 \end{figure*}

\begin{figure*}
 \begin{center}
 \includegraphics[width=8cm]{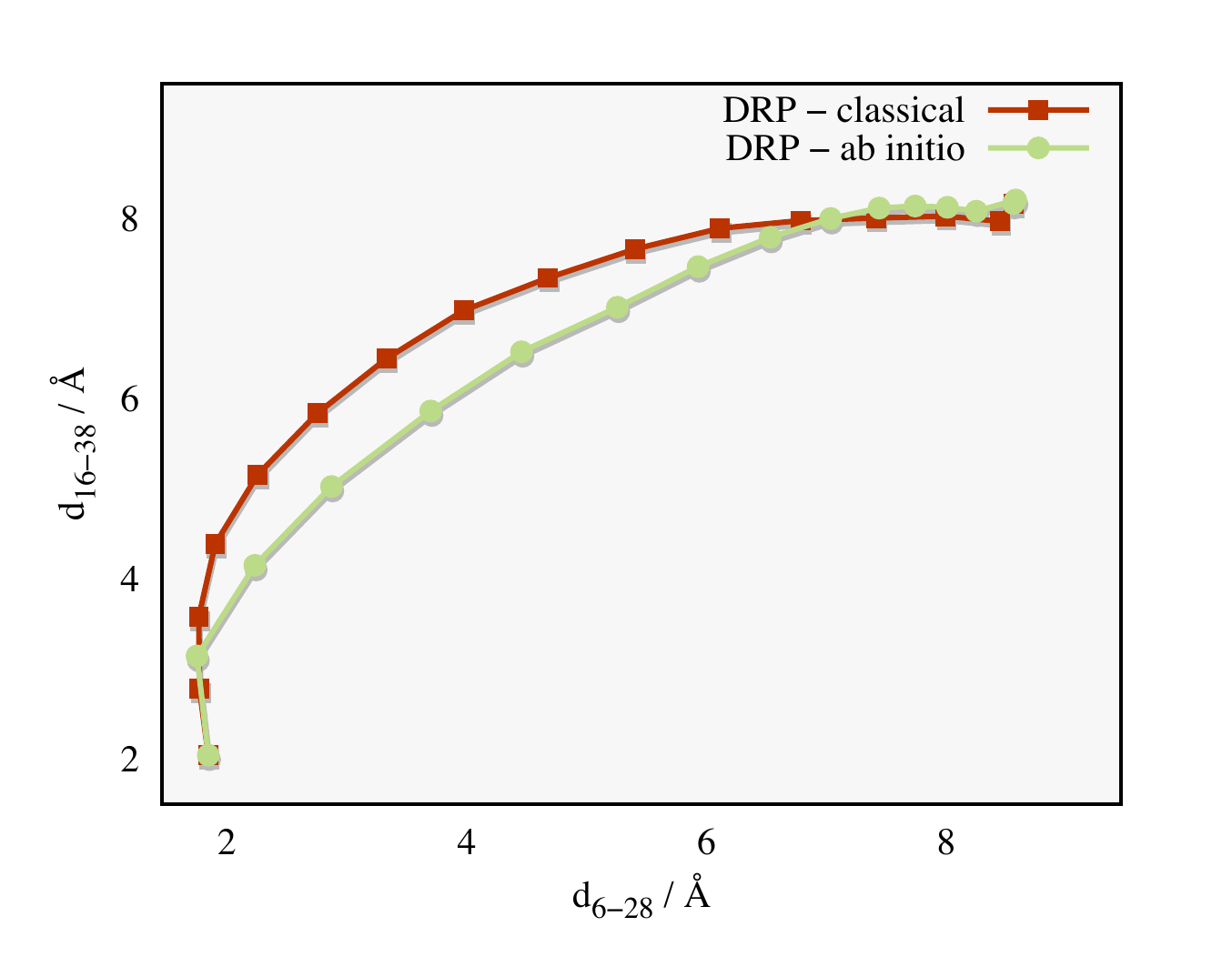}
 \end{center}
 \caption{Comparison of the dominant reaction pathways obtained from classical and \emph{ab-initio} DRP simulations, projected onto the plane selected by the $d_{6-28}$ and
 $d_{16-38}$ order parameters}
 \label{hbonds}
 \end{figure*}

In Fig. \ref{orderparameters} we compare the evolution of these quantities, as a function of the reaction coordinate $l$  obtained from classical and \emph{ab-initio} simulations. We also plot the initial high-temperature MD path and the path obtained at the end of the preliminary  relaxation based on the NEB algorithm --- cfr. the discussion in the section "Model"---.
In Fig. \ref{hbonds} we show the same dominant paths, projected onto the planes selected by the distances $d_{6-28}$ and $d_{16-38}$ involved in the formation of hydrogen bonds.

Some comments on these results are in order. First of all, these plots clearly show that  the FIRE algorithm allows to sizeably move away from the initial trial path, irrespective of the order parameter used to characterize the transition.
Secondly, we observe that the dominant paths obtained from MM and \textit{ab-initio} simulations agree at an almost quantitative level.  
In particular, both calculations predict that the contact between the $C-6$  and the $H-28$ atoms is formed before the contact between the $O-16$ and the $H-38$ atoms. However, in the MM calculation, the formation of the second contact occurs at a slightly later stage of the reaction than in the quantum calculation.   
We emphasize that the observed good agreement between classical and quantum calculations is not due to the insufficient exploration of the path space. In fact, such dominant paths are qualitatively different from those obtained in the last stage of the preliminary NEB minimization. 
Thus, these results clearly imply that the AMBER--99 classical force field  provides a rather accurate description of the dynamics, even in non-equilibrium conditions.

\begin{figure}[t]
\begin{center}
\includegraphics[width=8cm]{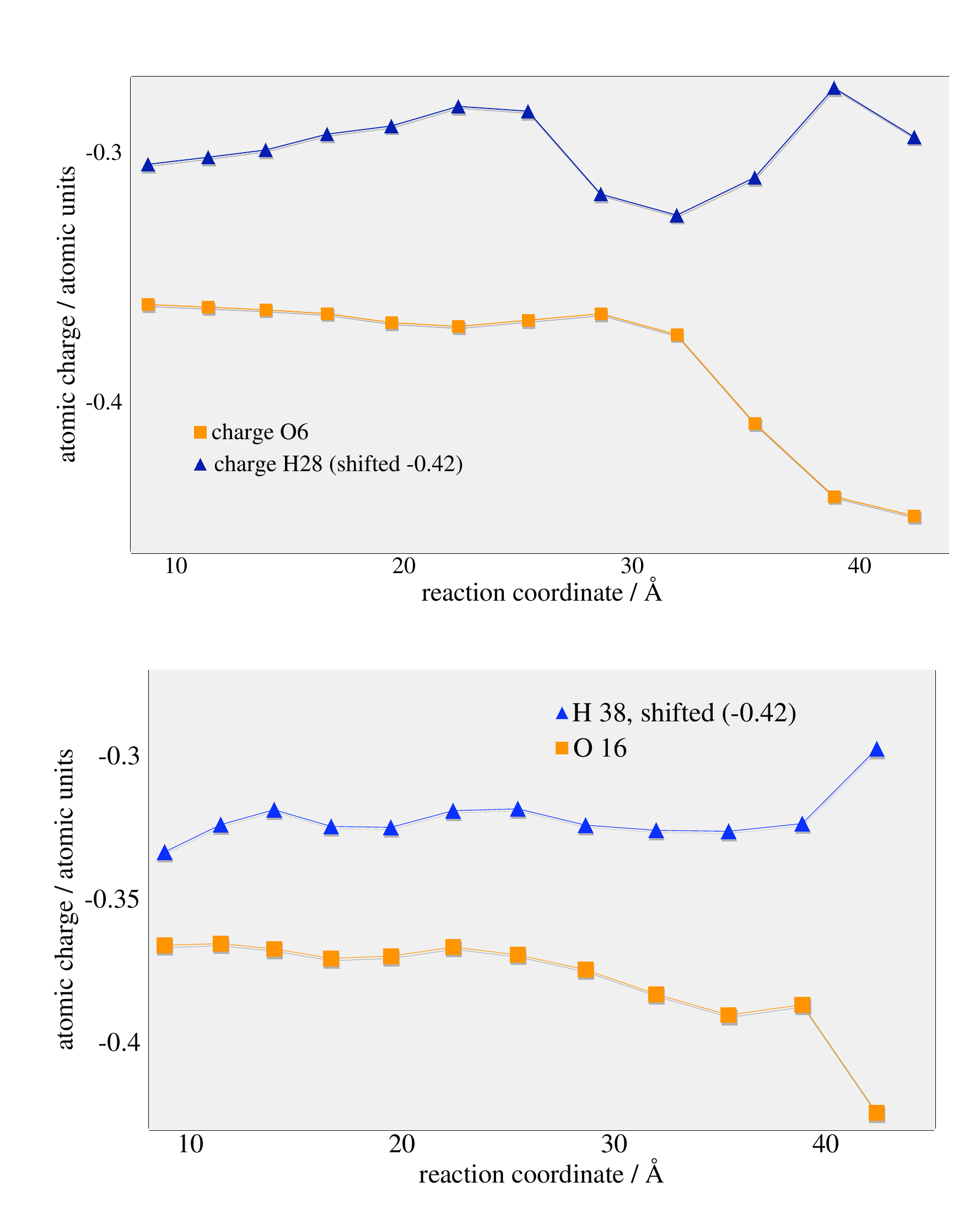}
\end{center}
 \caption{Evolution of the partial charges of the $O-6$ and $H-28$ atoms (upper panel) and of the $O-16$ and $H-38$ atoms (lower panel), along the dominant reaction pathway.
 In both figures, the partial charge of the $H$ atoms has been shifted by $-0.42$ atomic units for seek of graphical clarity. }
 \label{partialcharges}
  \end{figure}

The quantum  calculation provides additional physical information, not accessible by means of MM simulations.
For example, in Fig. \ref{partialcharges} we plot the evolution of the partial charges of $O-6$ and $H-28$ and of $O-16$ and $H-38$ during the reaction. We recall that these pairs of atoms form hydrogen bonds in the final helix state. By cross-correlating this information with the evolution of the interatomic distance ---cfr. Fig. \ref{orderparameters}--- we can infer that a significant modification of the electronic structure of the $O-6$ and $H-28$  atoms sets in when they are separated by a distance of the order 3~\AA.
Interestingly, the electronic structure of the  $O-16$ and $H-38$ begin to change when these two atoms are separated by a much larger distance, of the order of 4~\AA. 
These results exhibit an example of the fact that the validity  of some of the assumptions involved in MM calculations ---such as  the invariance of  partial charge---may depend  on the detail of the chemical environment in which the bond is formed and on the specific non-equilibrium dynamics of the reaction.  

Using Eq. (\ref{time}) it is in principle possible to obtain information about the dynamics, i.e. to compute the time  at which each of the configurations of the dominant path is 
visited during the course of the transition. In particular, assuming a  diffusion constant $D=2~\cdot~10^{-2}$~\AA$^2$~ps$^{-1}$  for all atoms, as in \cite{DRPtest1},  the total transition path times in the classical and quantum calculations are found to be $t_{classical} =  1.4$~ps and $t_{ab-initio} = 2. 3$~ps,  respectively. We note, however, that these numbers should be taken with  care, because in the present exploratory calculation we used only $N_s=16$ 
path discretization steps, and Eq. (\ref{time}) is known to be quite sensitive to discretization errors.

\section{Discussion}
In this work, we have used the DRP method to perform the first  simulation of the folding reaction of a peptide chain, based on a \emph{ab-initio} quantum-mechanical approach.  We have calculated the most statistically probable reaction pathway  connecting an initial coil configuration and a  final helix configuration, assuming the over-damped Langevin dynamics.

By comparing the results of the \emph{ab-initio} simulation in the PM3 scheme with MM simulations with the  AMBER--99 force field, we argue that MM approaches provide a quite reliable description even for out-of-equilibrium dynamics. We also studied the evolution of the partial charges involved in the formation of two hydrogen bonds stabilizing the helix, and found that the dynamics of these observable depends on the chemical environment.

The similarity of the quantum and classical paths suggests a  ``perturbative'' approach to performing \emph{ab-initio} DRP simulations at a much smaller computational cost.  Indeed, our results show that the minimum of the classical HJ functional lies in the vicinity of the minimum of the \emph{ab-initio} HJ functional. A first approximate dominant pathway can be calculated using the classical DRP approach, and used as a starting point for a further local minimization, in the \emph{ab-initio} framework. We expect that a few quantum mechanical minimization steps should be enough to reach convergence. 

We note that in order to characterize the folding mechanism, it should be taken into account that the structure of the folding pathways may significantly depend on the specific initial condition from which the transition is initiated. Hence, an exhaustive study of the folding dynamics of this system requires a statistical analysis of an \emph{ensemble} of folding trajectories, corresponding to different initial conditions, as discussed in \cite{DRPtest2}~\footnote{We thank R.Elber for an important discussion on this point}. 

We conclude this paper by discussing the computational cost of performing a similar calculation for a larger system, for instance a 15-residue $\beta$-hairpin in implicit solvent. The scaling of MOPAC energy calculation with the number of atoms can be made linear for large molecules by introducing a cut-off for matrix elements of the Hamiltonian between orbitals on atoms beyond a certain distance. Using a discretization of the path in N$_s$=100 slices, we estimate that a \textit{ab-initio} DRP calculation could be carried out using approximately 500,000 total CPU hours per trajectory.

While this is a substantial amount of computing time, we point out that it can be achieved on existing large computing facilities.

\begin{acknowledgments} S. a Beccara, P. Faccioli, F. Pederiva and M. Sega 
are members of the Interdisciplinary Laboratory for
Computational Sciences (LISC), a joint venture of the University of Trento
and Fondazione Bruno Kessler.  F. P. thanks the Quantum Simulations Group at Lawrence
Livermore National Laboratory for providing access to computing facilities.
Part of the calculations was performed on the WIGLAF cluster at the University of Trento.
\end{acknowledgments}

\newpage


\end{document}